\documentclass[final,3p,times,twocolumn]{elsarticle}

\usepackage{amssymb}
\usepackage{multirow}
\usepackage{amsmath}
\usepackage{url}

\usepackage[colorlinks=true, citecolor=blue, urlcolor=blue]{hyperref}

\usepackage{lineno}

\DeclareMathAlphabet{\mathnormal}{OT1}{cmr}{m}{n}

\biboptions{numbers,sort&compress}

\journal{Astroparticle Physics}

\begin{document}

\begin{frontmatter}



\title{Antihelium-3 Sensitivity for the GRAMS Experiment}

\author[Northeastern]{J. Zeng} 
\ead{zeng.jia@northeastern.edu}
\author[Northeastern]{T. Aramaki} 
\author[Waseda,JAXA]{K.~Aoyama}
\author[Tokyo]{S.~Arai}
\author[Waseda]{S.~Arai}
\author[UTA]{J.~Asaadi}
\author[Tokyo]{A.~Bamba}
\author[GSFC]{N.~Cannady}
\author[Yale]{P.~Coppi}
\author[GSFC]{G.~De Nolfo}
\author[WU]{M.~Errando}
\author[ORNL]{L.~Fabris}
\author[Osaka]{T.~Fujiwara}
\author[Hiroshima]{Y.~Fukazawa}
\author[GSFC]{P.~Ghosh}
\author[Tokyo]{K.~Hagino}
\author[Osaka]{T.~Hakamata}
\author[Yokohama]{N.~Hiroshima}
\author[Tokyo]{M.~Ichihashi}
\author[RIKEN]{Y.~Ichinohe}
\author[Osaka,iTHEMS,IPMU]{Y.~Inoue}
\author[Waseda]{K.~Ishikawa}
\author[Osaka]{K.~Ishiwata}
\author[Tokyo]{T.~Iwata}
\author[Columbia]{G.~Karagiorgi}
\author[Tokyo]{T.~Kato}
\author[Osaka]{H.~Kawamura}
\author[CAS,TIANFU]{D.~Khangulyan}
\author[GSFC]{J.~Krizmanic}
\author[Northeastern]{J.~LeyVa}
\author[Columbia]{A.~Malige}
\author[GSFC]{J.G.~Mitchell}
\author[GSFC]{J.W.~Mitchell}
\author[Barnard]{R.~Mukherjee}
\author[Waseda]{R.~Nakajima}
\author[Nagoya]{K.~Nakazawa}
\author[Osaka]{H.~Odaka}
\author[Nagoya]{K.~Okuma}
\author[Columbia]{K.~Perez}
\author[Northeastern]{N.~Poudyal}
\author[Columbia]{I.~Safa}
\author[Chicago]{K.~Sakai}
\author[GSFC]{M.~Sasaki}
\author[Columbia]{W.~Seligman}
\author[Columbia]{J.~Sensenig}
\author[Osaka]{K.~Shirahama}
\author[Kanagawa]{T.~Shiraishi}
\author[Howard]{S.~Smith}
\author[Hiroshima]{Y.~Suda}
\author[Northeastern]{A.~Suraj}
\author[Hiroshima]{H.~Takahashi}
\author[Tokyo]{S.~Takashima}
\author[Tokyo,JAXA]{T.~Tamba}
\author[Waseda]{M.~Tanaka}
\author[Columbia]{S.~Tandon}
\author[Osaka]{R.~Tatsumi}
\author[UCB]{J.~Tomsick}
\author[Kanagawa]{N.~Tsuji}
\author[TokyoScience]{Y.~Uchida}
\author[Waseda]{Y.~Utsumi}
\author[Howard]{S.~Watanabe}
\author[Waseda]{Y.~Yano}
\author[NDMC]{K.~Yawata}
\author[Wurzburg]{H.~Yoneda}
\author[Waseda]{K.~Yorita}
\author[Osaka]{M.~Yoshimoto}

\address[Northeastern]{Department of Physics, Northeastern University, 360 Huntington Avenue, Boston, MA 02115, USA}
\address[Waseda]{Department of Physics, Waseda University, 3-4-1 Okubo, Shinjuku-ku, Tokyo 169-8555, Japan}
\address[JAXA]{Japan Aerospace Exploration Agency (JAXA), 3-1-1 Yoshinodai, Chuo-ku, Sagamihara City, Kanagawa 252-5210, Japan}
\address[Tokyo]{Department of Physics, University of Tokyo, Tokyo 113-0033, Japan}
\address[UTA]{Department of Physics, University Texas Arlington, 701 South Nedderman Drive, Arlington, TX 76019, USA}
\address[GSFC]{NASA Goddard Space Flight Center, 8800 Greenbelt Road, Greenbelt, MD 20771, USA}
\address[Yale]{Department of Astronomy, Yale University, P.O. Box 208101 New Haven, CT 06520-8101, USA}
\address[WU]{Department of Physics, Washington University at St. Louis, One Brookings Drive, St. Louis, MO 63130-4899, USA}
\address[ORNL]{Oak Ridge National Laboratory, 5200, 1 Bethel Valley Rd, Oak Ridge, TN 37830, USA}
\address[Osaka]{Department of Earth and Space Science, Graduate School of Science, Osaka University, 1-1 Machikaneyama-cho, Toyonaka, Osaka 560-0043, Japan}
\address[Hiroshima]{Department of Physics, Graduate School of Advanced Science and Engineering, Hiroshima University, 1-3-2, Kagamiyama, Higashi Hiroshima-shi, Hiroshima 739-0046, Japan}
\address[Yokohama]{Department of Physics, Faculty of Engineering Science, Yokohama National University, Yokohama 240-8501, Japan}
\address[RIKEN]{RIKEN Nishina Center, Hirosawa 2-1, Wako-shi, Saitama 351-0198, Japan}
\address[iTHEMS]{Interdisciplinary Theoretical \& Mathematical Science Program (iTHEMS), RIKEN, 2-1 Hirosawa, 351-0198, Japan}
\address[IPMU]{Kavli Institute for the Physics and Mathematics of the Universe (WPI), UTIAS, The University of Tokyo, 5-1-5 Kashiwanoha, Kashiwa, Chiba 277-8583, Japan}
\address[Columbia]{Department of Physics, Columbia University, New York, NY 10027, USA}
\address[CAS]{Key Laboratory of Particle Astrophyics, Institute of High Energy Physics, Chinese Academy of Sciences, 100049 Beijing, China}
\address[TIANFU]{Tianfu Cosmic Ray Research Center, 610000 Chengdu, Sichuan, China}
\address[Barnard]{Department of Physics and Astronomy, Barnard College, 3009 Broadway, New York, NY 10027, USA}
\address[Chicago]{Enrico Fermi Institute, The University of Chicago, 5640 South Ellis Ave Chicago IL 60637, USA}
\address[Nagoya]{Department of Physics, Nagoya University, Furo-cho, Chikusa-ku, Nagoya, Aichi 464-8601, Japan}
\address[Kanagawa]{Faculty of Science, Kanagawa University, 3-27-1, Rokkakubashi, Kanagawa-ku, Yokohama-shi, Kanagawa 221-0802, Japan}
\address[Howard]{Department of Mechanical Engineering, Howard University, 2400 6th St NW, Washington, DC 20059, USA}
\address[UCB]{University of California Berkeley Space Sciences Laboratory, University Avenue and, Oxford St, Berkeley, CA 94720, USA}
\address[TokyoScience]{Department of Physics, Faculty of Science and Technology, Tokyo University of Science, 2641 Yamazaki, Noda, Chiba 278-8510, Japan}
\address[NDMC]{Department of Medical Education, National Defense Medical College, 3-2 Namiki, Tokorozawa, Saitama 359-8513, Japan}
\address[Wurzburg]{Julius-Maximilians-Universit\"{a}t W\"{u}rzburg, Fakult\"{a}t f\"{u}r Physik und Astronomie, Institut f\"{u}r Theoretische Physik und Astrophysik, Lehrstuhl f\"{u}r
Astronomie, Emil-Fischer-Str. 31, D-97074 Würzburg, Germany, Sanderring 2, 97070 W\"{u}rzburg, Germany}





\begin{abstract}

The Gamma-Ray and AntiMatter Survey (GRAMS) is a next-generation balloon/satellite mission utilizing a Liquid Argon Time Projection Chamber (LArTPC) detector to measure both MeV gamma rays and antinuclei produced by dark matter annihilation or decay. The GRAMS can identify antihelium-3 events based on the measurements of X-rays and charged pions from the decay of the exotic atoms, Time of Flight (TOF), energy deposition, and stopping range. This paper shows the antihelium-3 sensitivity estimation using a GEANT4 Monte Carlo simulation. For the proposed long-duration balloon (LDB) flight program (35 days $ \times $ 3 flights) and future satellite mission (2-year observation / 10-year observation), the sensitivities become 1.47 $\times$ 10$^{-7}$ [m$^2$ s sr GeV/n]$^{-1}$ and 1.55 $\times$ 10$^{-9}$ [m$^2$ s sr GeV/n]$^{-1}$ / $3.10\times10^{-10}$ [m$^2$ s sr GeV/n]$^{-1}$, respectively. The results indicate that GRAMS can extensively investigate various dark matter models through the antihelium-3 measurements.

\end{abstract}


\end{frontmatter}


\section{Introduction}
\label{section: Introduction}

\subsection{Dark Matter}
\label{subsection:Dark Matter}

The Planck experiment provides evidence that 68\% of our universe comprises dark energy, 27\% is dark matter and 5\% is baryonic matter \cite{10.1051/0004-6361/201321591}. The existence of dark matter is supported by multiple astronomical observations including galaxy rotation curves and gravitational lensing in the Bullet Cluster, where two colliding galaxy clusters exhibit clear separation between their mass and baryonic components \cite{10.1046/j.1365-8711.2000.03075.x}. Despite the observational evidence, the nature of dark matter and its interactions with ordinary matter remain poorly understood. 

Several theoretical frameworks attempt to explain dark matter, with Weakly Interacting Massive Particles (WIMPs) being one of the leading candidates. Proposed WIMP candidates include neutralinos, right-handed sneutrinos, and right-handed neutrinos in extra dimension theories \cite{donato2000antideuterons, donato2008antideuteron,cerdeno2009right, cerdeno2014low,baer2005low}. Astrophysics experiments to detect dark matter, both directly and indirectly, are ongoing, including satellite missions like the Alpha Magnetic Spectrometer-02 (AMS-02) and the Fermi Gamma-ray Space Telescope (Fermi), as well as balloon-borne experiments such as the Balloon-borne Experiment with a Superconducting Spectrometer (BESS) and the General Antiparticle Spectrometer (GAPS) \cite{aguilar2002alpha, lubelsmeyer2011upgrade, ajima2000superconducting, hailey2006accelerator}.

\subsection{Antihelium-3 as Dark Matter Search}
\label{subsection:antihelium3 as Dark Matter Search}
AMS-02 has reported the detection of antihelium-like events, generating significant interest in the development of dedicated antihelium detection capabilities by current and future experiments \cite{aguilar2013first}. These observations are particularly intriguing as they may indicate antihelium production through dark matter annihilation processes since standard astrophysical production of antihelium nuclei via cosmic interactions is highly suppressed \cite{PhysRevD.99.023016, PhysRevD.102.063004, Kachelrie_2020}.

GRAMS represents a novel approach to indirect dark matter detection, specifically optimized for antinuclei measurements. GRAMS employs liquid argon as its detection medium, enabling the capture of antiparticles and the subsequent decay of their annihilation products. This detection strategy provides a unique window into potential dark matter signatures through antihelium-3 measurements. GRAMS is capable of measuring sub-GeV/n antiprotons, which will provide cross-checking with previous cosmic antiproton measurements as well as instrumental calibration to validate the antinuclei detection method. GRAMS will also provide low-energy antideuteron measurements with an extensive sensitivity of $\sim$10$^{-6}$ [m$^2$ s sr GeV/n]$^{-1}$ as described in the concept paper \cite{aramaki2020dual}. This work focuses on and presents the expected GRAMS sensitivity for antihelium-3 searches. 

\section{GRAMS Project Overview}
\label{section:GRAMS project overview}

GRAMS is a next-generation experiment that aims at detecting both astrophysical observations with MeV gamma rays and indirect dark matter searches with antimatter using a LArTPC (Liquid Argon Time Projection Chamber) detector \cite{aramaki2020dual}.

\begin{figure}[htbp]
\begin{center} 
\includegraphics*[width=7.5cm]{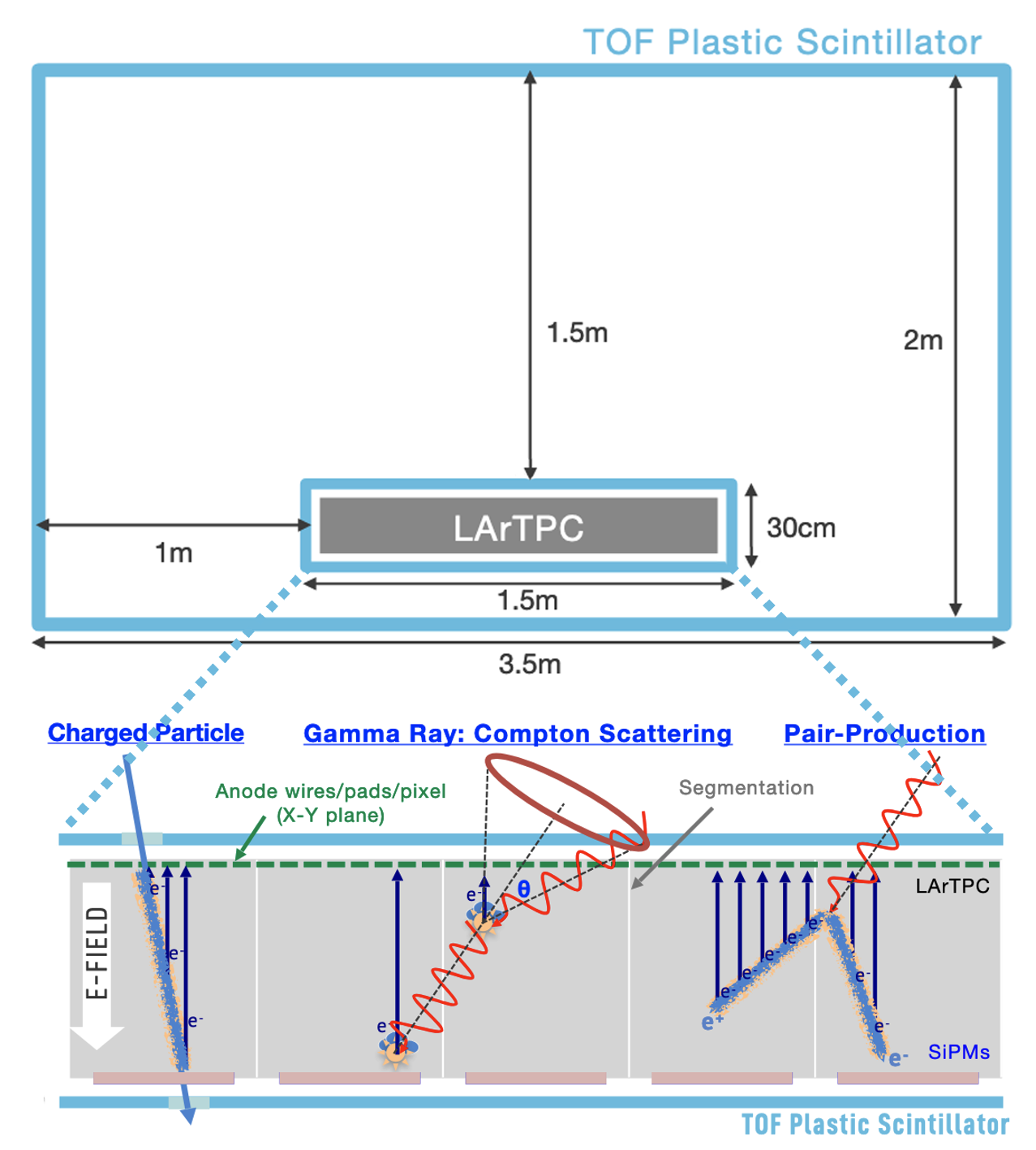}
\end{center}
\caption{GRAMS detection concept. The LArTPC is segmented into “cells” to minimize coincident background events. The bottom figure shows the charged-particle and gamma-ray interactions inside the detector.}
\label{schematic}
\end{figure}

\subsection{Detection Concept for Antiparticles}
\label{subsection: Detection Concept}

The GRAMS detection concept utilizes the combined signals from time-of-flight (TOF) and LArTPC systems to identify and reconstruct events. The TOF system will measure the velocity of the incoming particle and its energy deposition. The incoming charged antiparticles will slow down as they deposit energy, through ionization, in the LAr. The antiparticle will be combined with an argon nucleus, forming an exotic atom. The exotic atom in the excited state will de-excite, emitting Auger electrons and X-rays \cite{aramaki2013measurement}. The antiparticle will eventually be captured by the nucleus and produce annihilation products, including charged pions and protons. The number of pions and protons produced here will be related to the number of antinucleons, providing additional information to identify the incoming antiparticle (see Sec.~\ref{subsection: Separation}).

Since antiprotons will also form exotic atoms in the LAr detector and generate annihilation products at the stop position, they will contribute the majority of background events for antihelium-3 detection. However, an antihelium-3 nucleus will deposit more energy in the TOF system as it has a charge of -2e (see Sec.~\ref{subsubsection: energy deposition}). Moreover, since it consists of three antinucleons, the annihilation product profile will be different from an antiproton, where more protons and pions can be generated from the annihilation point.

\subsection{GRAMS Balloon Instrumental Design}
\label{subsection: Instrumental Design}

GRAMS's balloon flight will be the first large-scale LArTPC experiment targeting MeV gamma rays and antiparticles. The outer TOF system (3.6 m $\times$ 3.6 m $\times$ 2.0 m) and inner TOF system (1.5 m $\times$ 1.5 m $\times$ 0.3 m) will be a series of segmented plastic scintillators to measure the incoming particles' timing information. The LArTPC (1.4 m $\times$ 1.4 m $\times$ 0.2 m) surrounded by the inner TOF will be segmented into cells to reduce the coincident background events. The LArTPC will have an anode (cathode) plane at the top (bottom). The cathode plane will include a 2D electronic readout tile to measure the ionized electrons and determine the x and y positions for the event \cite{albert2018sensitivity}. The z position can be determined based on the electron's drifting time after the scintillation light is measured by silicon photomultipliers (SiPMs).

\subsection{GRAMS Current Status and Future}
\label{subsection: GRAMS Current Status}

Aiming to realize these concepts and design, we are currently testing a small-scale LArTPC detector, MicroGRAMS (10 cm $\times$ 10 cm $\times$ 10 cm), at Northeastern University to validate its detection performance. An engineering balloon flight (eGRAMS) was successfully launched in 2023 from the JAXA Taiki Aerospace Research Field, which verified the feasibility and functionality of the LArTPC operation under balloon flight conditions in space \cite{Nakajima:2024fgx}.

GRAMS has been funded by the NASA APRA program for the prototype balloon flight (pGRAMS) scheduled in the Spring of 2026, where we will deploy a prototype LArTPC detector, MiniGRAMS (30 cm $\times$ 30 cm $\times$ 20 cm) \cite{nasa_suborbital}. MiniGRAMS will also be used in the first science flight after pGRAMS, followed by a science balloon flight with a full-scale GRAMS LArTPC (140 cm $\times$ 140 cm $\times$ 20 cm). The GRAMS project will eventually be expanded to a satellite mission. With longer observation time and large-scale LArTPC, GRAMS could significantly improve the sensitivities for both MeV gamma-ray and antinuclei measurements \cite{aramaki2020dual}. In particular, this paper will describe how we evaluate the GRAMS sensitivity to antihelium-3 nuclei and show GRAMS could potentially probe various dark matter models.

\section{GRAMS Antihelium-3 Sensitivity Calculation}
\label{section: GRAMS antihelium3 sensitivity calculation}
\subsection{GRASP}
\label{subsection: GRASP}
\begin{figure}[htbp]
\begin{center} 
\includegraphics*[width=7.5cm]{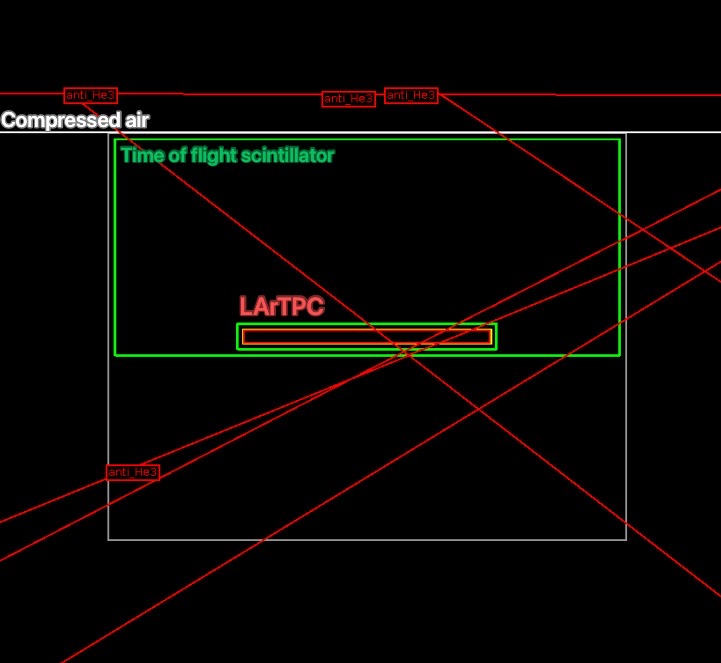}
\end{center}
\caption{Geant4 simulation with the GRAMS geometry for the GRASP estimation. We generated antiparticles on a 20 m $\times$ 20 m plane located above a thin layer of compressed air and the GRAMS payload. Antiparticles go through the compressed air to simulate the atmospheric effect. Two green boxes are the outer and inner TOF paddles, and the red cube inside the inner TOF box is the LArTPC detector (140 cm $\times$ 140 cm $\times$ 20 cm). The red lines are the example tracks of the generated antihelium-3 nuclei.}
\label{schematic}
\end{figure}

We evaluated the sensitivity of GRAMS for antihelium-3 nuclei using GEANT4 Monte Carlo simulations with QGSP\_BERT physics list \cite{AGOSTINELLI2003250, APOSTOLAKIS2009859}. Fig.~\ref{schematic} shows the simulation setup. Antiparticles are isotropically generated from a 20 m $\times$ 20 m plane at the top of the atmosphere (TOA) with energy up to 1000 MeV/n and will propagate downward through a 3.9 cm compressed air layer with a 1 g/cm$^2$ density. The antiparticles will annihilate or lose energy, providing a realistic estimate of the antiparticle flux reaching the detector at the flight altitude of around 37 km. We define GRASP below to quantify the antiparticle stopping or inflight annihilation efficiency inside the detector. 

\begin{equation}
    \Gamma_i = \pi\cdot A\cdot\frac{N_{i}}{N_t}
\end{equation}
Here, $ A $ represents the area for the antiparticle generation, which is a 20 m $\times$ 20 m source plane, $ N_t $ is the number of total antiparticles generated in the area (50 million of antiproton and antihelium-3 events), and $ N_i $ is the number of antiparticles stopped or inflight annihilated inside the LArTPC detector. GRASP was calculated to characterize the detector's response as a function of initial energy at TOA. 

\begin{figure}[htbp]
\begin{center} 
\includegraphics*[width=7.5cm]{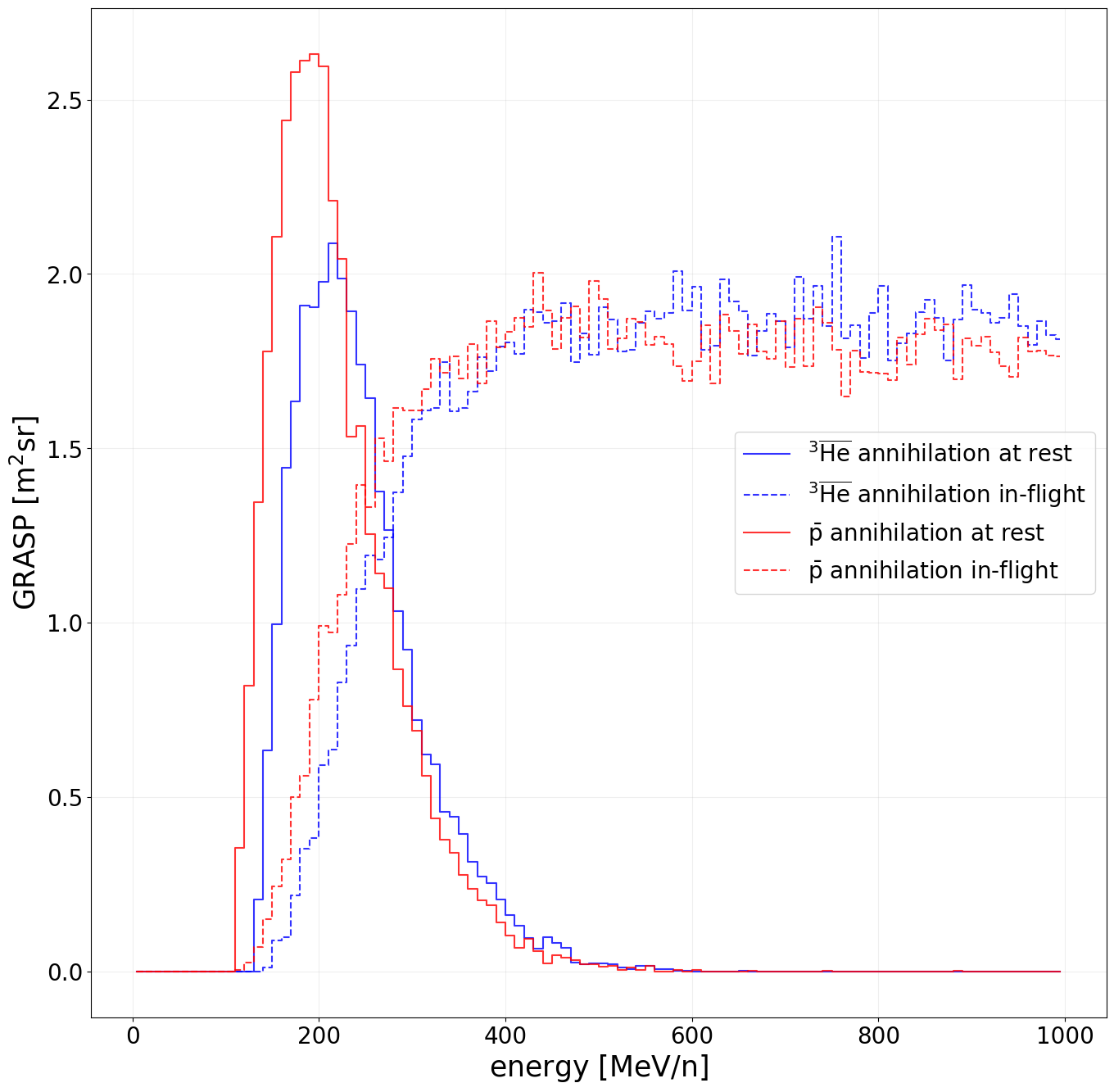}
\end{center}
\caption{GRASP simulation results with 50 million antiparticles generated each. Solid lines show annihilation at rest events, while dashed lines show annihilation in-flight events.}
\label{GRASP}
\end{figure}

Fig.~\ref{GRASP} shows the GRASP simulation results. Antiparticles with TOA energy between 100 MeV/n and 500 MeV/n can stop in the LArTPC, while inflight annihilation events dominate for antiparticles with energy greater than 300 MeV/n. The simulation result using the same GEANT4 framework showed a good match of the event rate at the balloon altitude during the engineering flight (eGRAMS) \cite{Nakajima:2024fgx}.

\subsection{Particle Identification Technique}
\label{subsection: Separation}

The main background for antihelium-3 nuclei measurements would be antiprotons, as mentioned in sec.~\ref{subsection: Detection Concept}. However, they can be identified and distinguished from antihelium-3 events based on particle identification techniques with the TOF profile for both timing and energy deposition, the number of annihilation products, and energies of the atomic X-rays from exotic atoms.  

\subsubsection{Energy deposition}
\label{subsubsection: energy deposition}

\begin{figure}[htbp]
\begin{center} 
\includegraphics*[width=7.5cm]{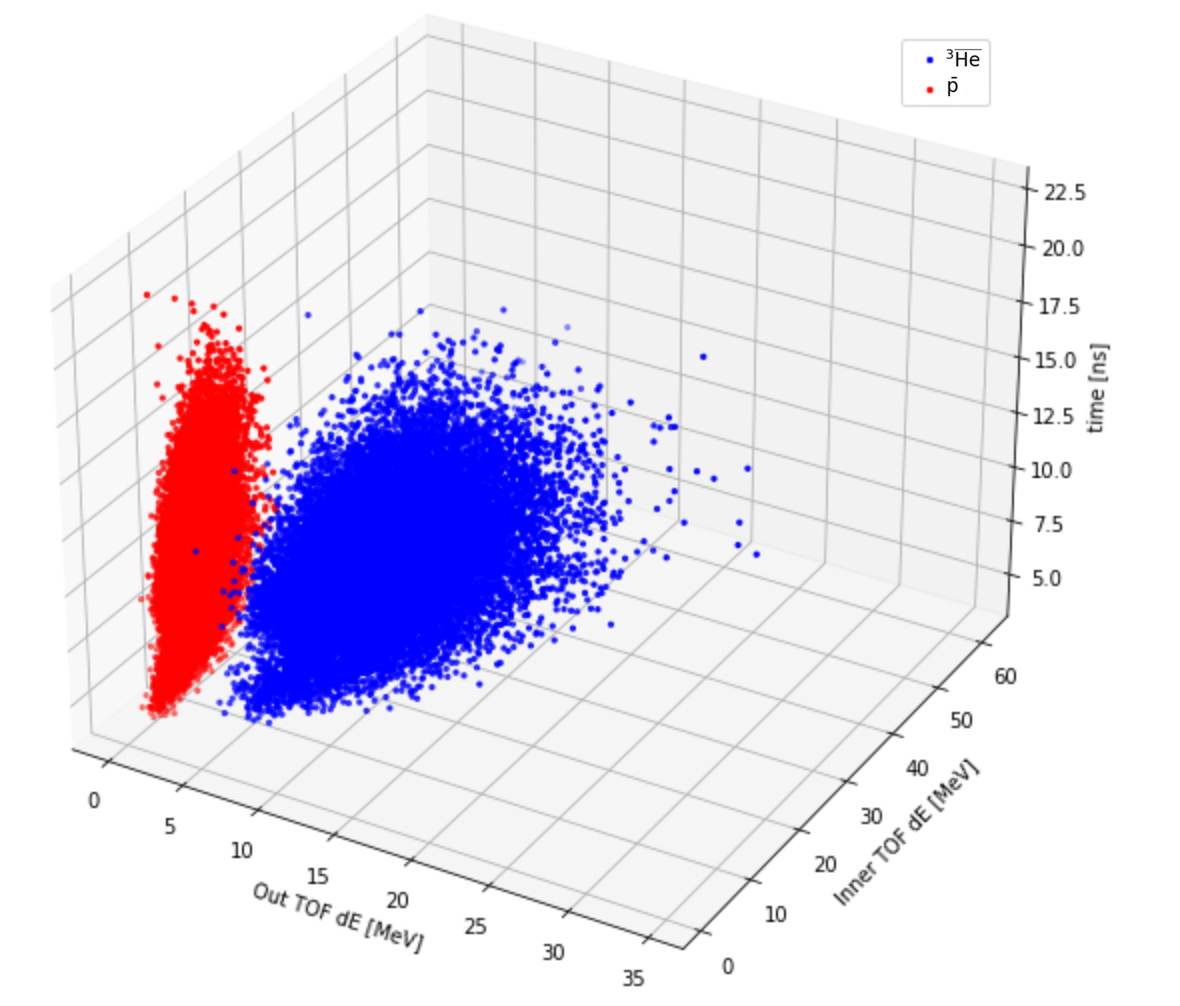}
\end{center}
\caption{The x-axis is the energy deposition in the outer TOF paddle, and the y-axis is the energy deposition in the inner paddle. The z-axis is the timing difference between the outer and inner paddle hits, which is related to the kinetic energy of the incoming charge particles. We can clearly see antiproton and antihelium-3 event clusters well separated from each other.}
\label{3D}
\end{figure}

Since antihelium-3 nuclei have a charge of -2e and roughly three times the antiproton mass, they will deposit more energy inside the TOF plastic scintillator with the same velocity measured by the TOF system compared to antiprotons. Geant4 simulations were conducted to evaluate the energy depositions in the outer and inner plastic scintillator paddles and the velocity based on the timing between the inner and outer TOF paddles. Here, energy resolution was assumed to be 16\% (14\%) for a charge of $\pm$2e ($\pm$1e) particles, and the timing resolution was considered to be 0.4 ns, as measured in other experiments, as well as the preliminary bench test in lab \cite{bindi2010scintillator}. Based on hit locations, timing, and accumulated energy deposition in the inner and outer TOF paddles, we obtained clusters for antiproton and antihelium-3 events in a 3D plot (see Fig.~\ref{3D}).

\begin{figure}
\begin{center} 
\includegraphics*[width=5cm]{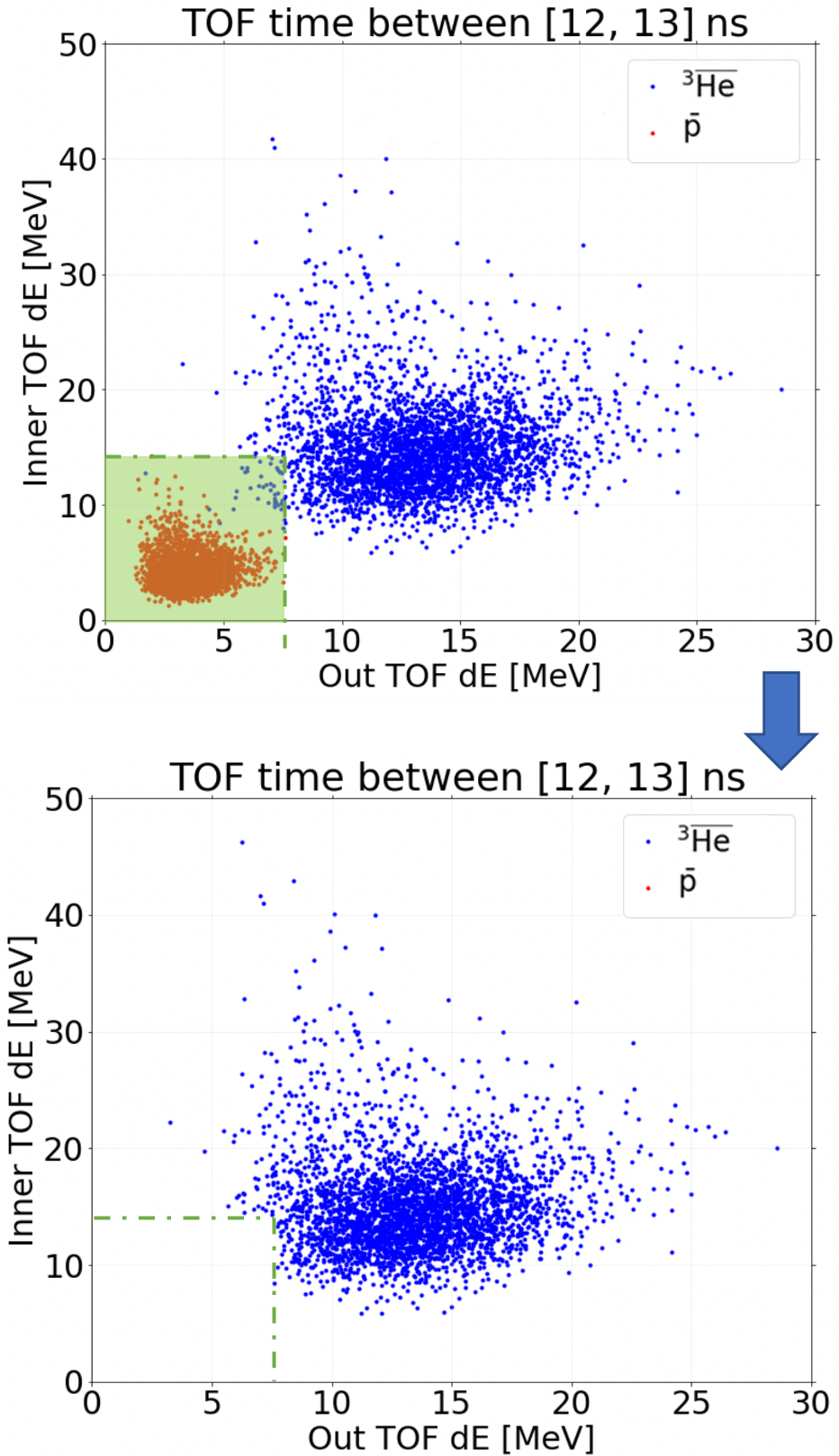}
\end{center}
\caption{An example of the sliced 2D plot for the TOF timing between 12 ns and 13 ns. We removed the light green shaded area at the bottom left to keep the rest of the region accepted. For this particular timing range, 98.6 \% of antihelium-3 events can be accepted while rejecting all antiproton events.}
\label{cut}
\end{figure}

\begin{figure}[htbp]
\begin{center} 
\includegraphics*[width=7.5
cm]{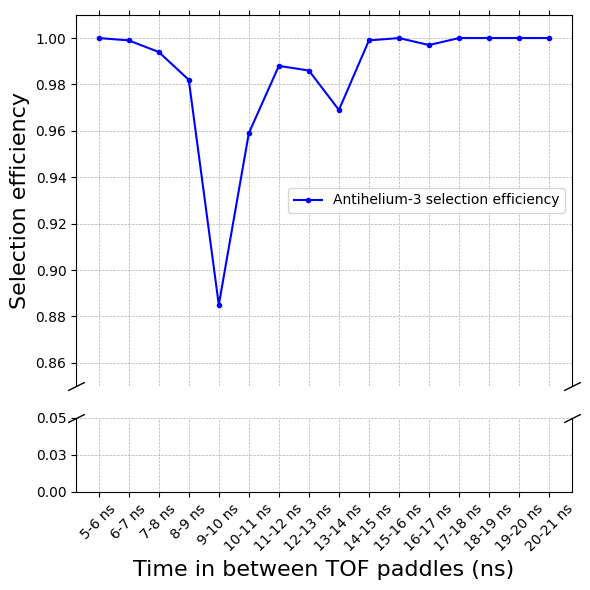}
\end{center}
\caption{Selection efficiency for antihelium-3 events under each one ns time window. Combined all simulation data gives an antihelium-3 nuclei selection cut
efficiency $\varepsilon_{TOF}^{^3\overline{He}}=0.965$, while we reject all antiprotons and provide an upper limit on the antiproton cut efficiency $\varepsilon_{TOF}^{\bar{p}}=0.000142$ (99\% C.L)}
\label{cut efficiency}
\end{figure}

To evaluate antihelium-3 selection cuts, we sliced the 3D plot with a time window of 1 ns ($\pm$0.5 ns) and made it into a 2D plot, energy depositions in the inner and outer TOF paddles. We applied the antihelium-3 selection cuts by drawing a box to completely cover the antiproton cluster (see an example in Fig.~\ref{cut}). With this cut method, we established a stair-shaped separation plane in the 3D plot. Fig.~\ref{cut efficiency} also shows the acceptance rate for antihelium-3 events with the applied cut, approximately 96.5 \%, while rejecting all simulated antiproton events. This gives an upper limit on the antiproton cut efficiency $\varepsilon_{TOF}^{\bar{p}}=0.000142$ (99\% C.L). Note that, aside from energy depositions inside TOF paddles, energy deposition inside the LArTPC detector can also be used to reject antiprotons, considering the charge and mass difference between antiprotons and antihelium-3 nuclei. 


\subsubsection{Charged pion multiplicity}
\label{subsubsection: Charged pion multiplicity}
When antiparticles annihilate inside the LArTPC, they will generate pions. The number of charged/neutral pions and its standard deviation ($\sigma$) for the antiproton-nucleon annihilation can be estimated as follows \cite{cugnon1989antiproton, cugnon1992antideuteron}.

\begin{equation}\label{equ: pion main}
    \langle M^p_{\pi^{\pm, 0}}\rangle=2.65+3.65\ln{\sqrt{s}}
\end{equation}

\begin{equation}\label{equ: pion sigma}
    \frac{\sigma^2}{\langle M^p_{\pi^{\pm, 0}}\rangle}=0.174(\sqrt{s})^{0.40}
\end{equation}

\begin{figure}[htbp]
\begin{center} 
\includegraphics*[width=7.5
cm]{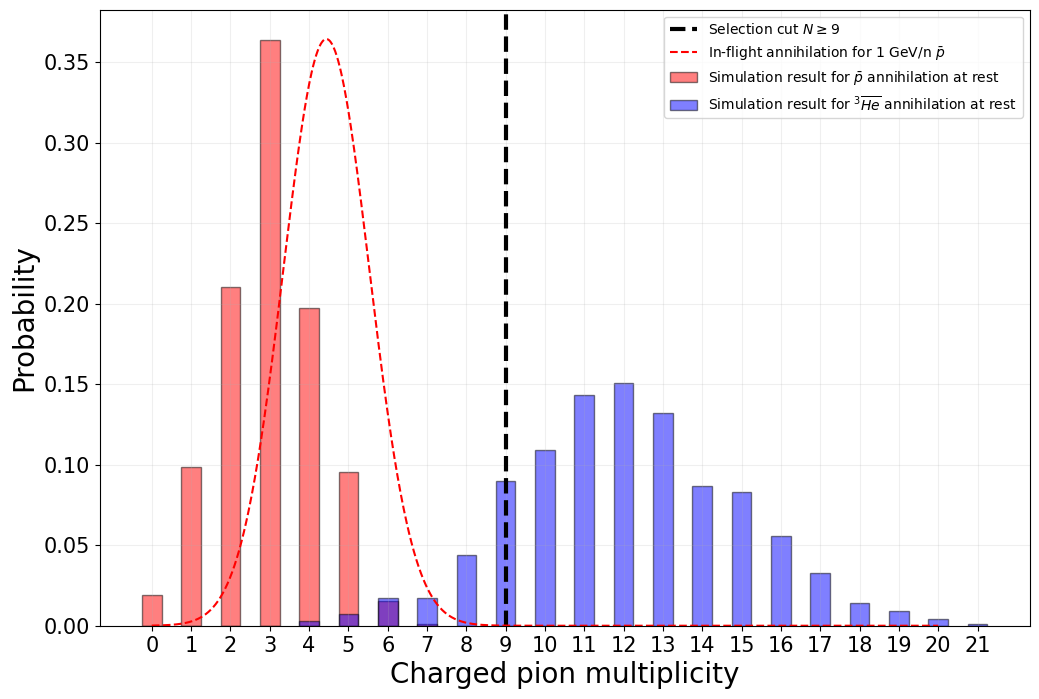}
\end{center}
\caption{GEANT4 simulation results of charged pion multiplicities for antiproton and antihelium-3 events are shown in red and blue bars. We could apply a cut of $N\geq9$ (vertical black dashed line) to reject antiproton events while still keeping $91.2\%$ of antihelium-3 events. The dashed red line shows the model prediction of the pion multiplicity for 1 GeV/n antiproton in-flight annihilation events, indicating the selection cut of $N\geq9$ is still valid with the antiproton selection efficiency of $\varepsilon_\pi^{\bar{p}}=1.25\times10^{-5}$.}
\label{Simulated Charged pion multiplicity}
\end{figure}

Here, $ \langle M^p_{\pi^{\pm, 0}}\rangle $ is the average number of charged and neutral pions, and $ \sqrt{s} $ is the center of mass energy in GeV \cite{aramaki2016antideuteron}. There are two models to analyze how antinucleons inside antihelium-3 nuclei interact with nuclei \cite{cugnon1992antideuteron}. We assume an even number of charged and neutral pions are generated here:

\begin{equation}
    \langle M_{\pi^+}\rangle = \langle M_{\pi^-}\rangle = \langle M_{\pi^0}\rangle
\end{equation}
Since neutral pions quickly decay into two high-energy gamma rays that can easily escape from the LArTPC, we will only consider charged pion multiplicity.
Fig.~\ref{Simulated Charged pion multiplicity} shows the GEANT4 simulation results of pion multiplicities from stopped 100000 antiprotons and antihelium-3 nuclei events. The antiproton result fits well with the theoretical model \cite{cugnon1989antiproton, cugnon1992antideuteron}. Here, we could apply a selection cut at 9 charged pions, providing below $10^{-5}$ of selection efficiency ($\varepsilon_\pi^{\bar{p}}$) for antiprotons while $\varepsilon_\pi^{^3\overline{He}}=0.912 $ for antihelium-3 events.

We used the equations (\ref{equ: pion main}) and (\ref{equ: pion sigma}) to estimate the pion multiplicity for antiproton in-flight annihilation events with the kinetic energy of 1 GeV/n (see the dashed red line in Fig.~\ref{Simulated Charged pion multiplicity}). Even for this case, we can easily separate antiproton and antihelium-3 events with the same selection cut, $N\geq9$ charged pions, providing the antiproton selection efficiency of $\varepsilon_\pi^{\bar{p}}=1.43\times10^{-5}$.

The INC model also predicts that protons and neutrons can be generated by three different processes: (1) direct emission from the interaction between the primordial pions and the nucleus, (2) pre-equilibrium emission (multifragmentation) from excited nucleons, and (3) nuclear evaporation \cite{chamberlain1955observation, aramaki2016antideuteron}. In future work, we may implement the proton multiplicity to enhance the antiproton rejection power.

\subsubsection{X-ray}
\label{subsubsection:X-Ray}

After forming exotic atoms, X-ray will be emitted during the de-excitation process. The energies of these atomic X-rays can be estimated based on the components of the exotic atoms as below:
\begin{equation}
    E_{X} = \left( zZ \right)^2 \frac{M^*}{m_e^*} R_H \left( \frac{1}{n^2_f} - \frac{1}{n^2_i} \right)
\end{equation}
Here, $z$ and $Z$ are the charges of the antiparticle and target atom, $M^*$ and $m_e^*$ are the reduced masses of an antiparticle in the exotic atom and an electron in the target atom, $R_H$ is the Rydberg constant, and $n_i$ and $n_f$ are the initial and final principal quantum numbers \cite{saffold2021cosmic}. This can result in different X-ray spectra for antiproton and antihelium-3 events, and we will implement this technique to enhance antiproton rejection power in future work. 

\subsection{Confidence Level and Sensitivity Calculation}
\label{subsection: Sensitivity and Confidence Level}

The main background events for GRAMS's antihelium-3 nuclei detection are antiprotons that can produce annihilation products in the LArTPC and the secondary antihelium-3 nuclei produced by cosmic-ray interactions. The antiproton and secondary antihelium-3 events, as well as the sensitivity of one antihelium-3 nuclei detection, $ S_{^3\overline{He}}$, can be estimated as follows.

\begin{equation}
    n_{\bar{p}}^{bkg}=\int F_{\bar{p}}(E)\Gamma_{\bar{p}}(E)T\varepsilon_g^{\bar{p}}dE\cdot\prod_i\varepsilon_i^{\bar{p}}
\end{equation}
\begin{equation}
    n_{^3\overline{He}}^{bkg}=\int F_{^3\overline{He}}^{sec}(E)\Gamma_{^3\overline{He}}(E)T\varepsilon_g^{^3\overline{He}}dE\cdot\prod_i\varepsilon_i^{^3\overline{He}}
\end{equation}
\begin{equation}
    S_{^3\overline{He}}=\frac{1}{\int \Gamma_{^3\overline{He}}(E)T\varepsilon_g^{^3\overline{He}}dE\cdot\prod_i\varepsilon_i^{^3\overline{He}}}
\end{equation}

\begin{table}[htbp]
\centering
\begin{tabular}{l|r}
\hline
&numerical value\\
\hline
$F_{\bar{p}}(E)$ & $10^{-2}$ [m$^2$ s sr GeV/n]$^{-1}$ \\
$F_{^3\overline{He}}^{sec}(E)$ &$5\times10^{-11}$ [m$^2$ s sr GeV/n]$^{-1}$ \\
A&$20\ m\times20\ m$ \\
$\int\Gamma_{\bar{p}}(E)dE$ & $1.77324\ m^2 sr \ GeV/n$ \\
$\int\Gamma_{^3\overline{He}}(E)dE$&$1.7017\ m^2 sr \ GeV/n$\\
T&105 days\\
$\varepsilon_g^{\bar{p}}$&0.7\\
$\varepsilon_g^{^3\overline{He}}$&0.5\\
$\varepsilon_{TOF}^{\bar{p}}$&0.000142\\
$\varepsilon_{TOF}^{^3\overline{He}}$&0.965\\
$\varepsilon_{\pi}^{\bar{p}}$&0.0000125\\
$\varepsilon_{\pi}^{^3\overline{He}}$&0.912\\
\hline
\end{tabular}
\caption{Sensitivity numerical values. Here, $ F_{\bar{p}}(E) $ and $ F_{^3\overline{He}}^{sec}(E) $ are the fluxes of antiprotons and secondary antihelium-3 nuclei at the top of the atmosphere\cite{PhysRevLett.105.121101, shukla2020large}. $\Gamma_{\bar{p}}(E)$ and $\Gamma_{^3\overline{He}}(E)$ are GRASPs for antiproton and antihelium-3 events, integrating through energy from Fig.~\ref{GRASP}. $ T $ will be the observation time during the level flight, and geomagnetic cutoff efficiency, $\varepsilon_g^{\bar{p}}$ for antiprotons and $\varepsilon_g^{^3\overline{He}}$ for antihelium-3 nuclei during the Antarctic flight \cite{vonDoetinchem:2017Id}. $ \varepsilon_i^{\bar{p}} $ is the acceptance for each antihelium-3 selection cut (charged pion multiplicity and TOF energy deposition). $\varepsilon_{TOF}^{\bar{p}}$ and $\varepsilon_{TOF}^{^3\overline{He}}$ are acceptance for antiproton and antihelium-3 cutnuclei TOF profile cut. while the pion multiplicity with $N\geq9$ gives selection cut acceptance $\varepsilon_{\pi}^{\bar{p}}$ and $\varepsilon_{\pi}^{^3\overline{He}}$ for antiproton and antihelium-3 nuclei respectively. }
\label{Balloon table}
\end{table}

Based on values show in table.~\ref{Balloon table}, the corresponding antihelium-3 sensitivity would be $1.47\times10^{-7}$ [m$^2$ s sr GeV/n]$^{-1}$ with the expected background events for antiproton (secondary antihelium-3) of $2.00\times10^{-4}$ ($3.4\times10^{-4}$). Fig.~\ref{antiHe3 sensitivity} shows the GRAMS sensitivity and antihelium-3 fluxes from dark matter models.

\subsection{GRAMS Satellite Mission}
\label{subsection: Sensitivity and Confidence Level}

\begin{figure}
\begin{center} 
\includegraphics*[width=7.5cm]{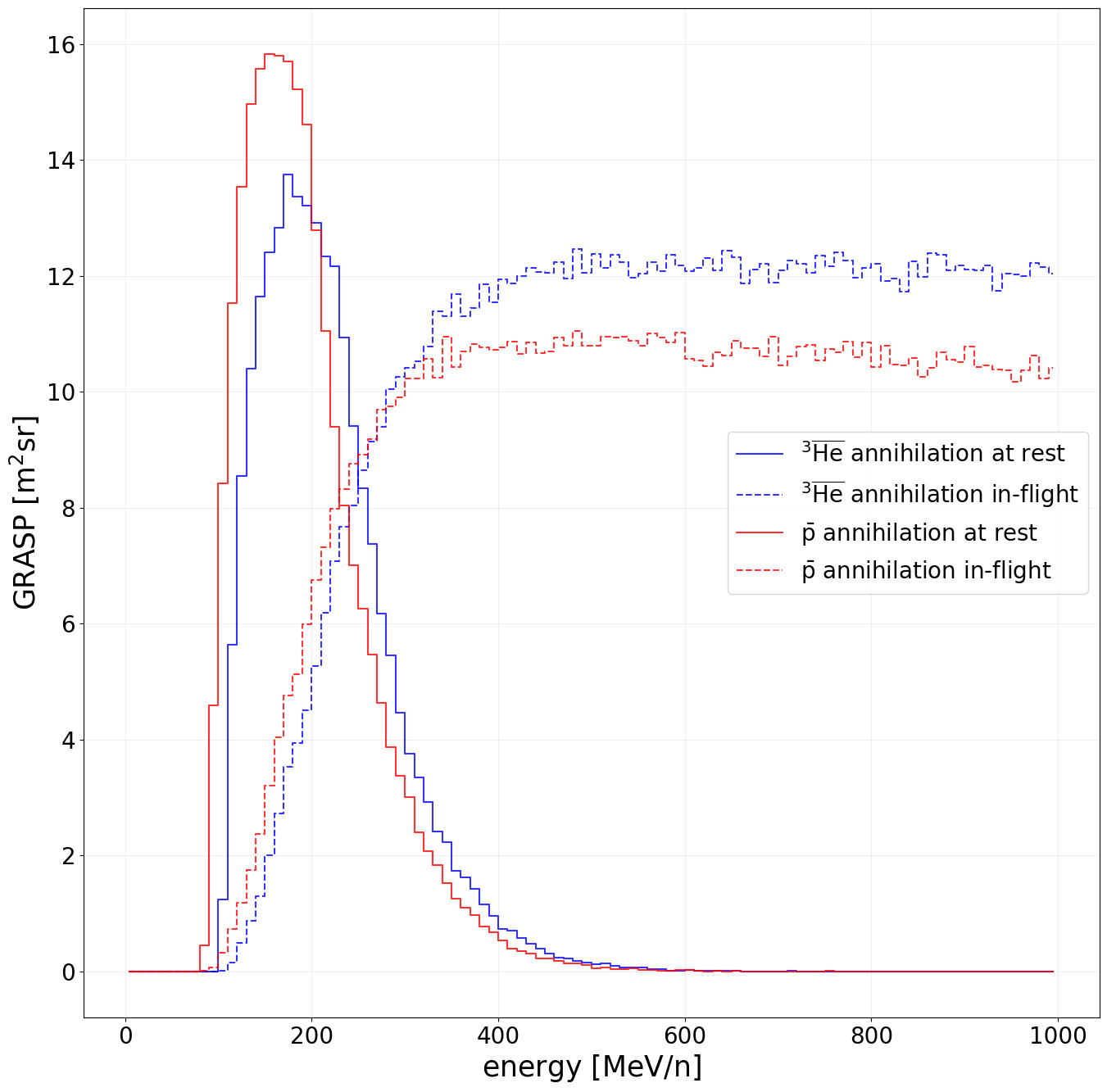}
\end{center}
\caption{GRASP estimation for the GRAMS satellite mission, with a 3.2 m $\times$ 3.2 m $\times$ 0.2 m LArTPC detector, considering the satellite fairing size \cite{7943833}. Solid lines show annihilation at rest events, while dashed lines show annihilation in-flight events.}
\label{GRASP for MAX design}
\end{figure}

\begin{figure}
\begin{center} 
\includegraphics*[width=7.5
cm]{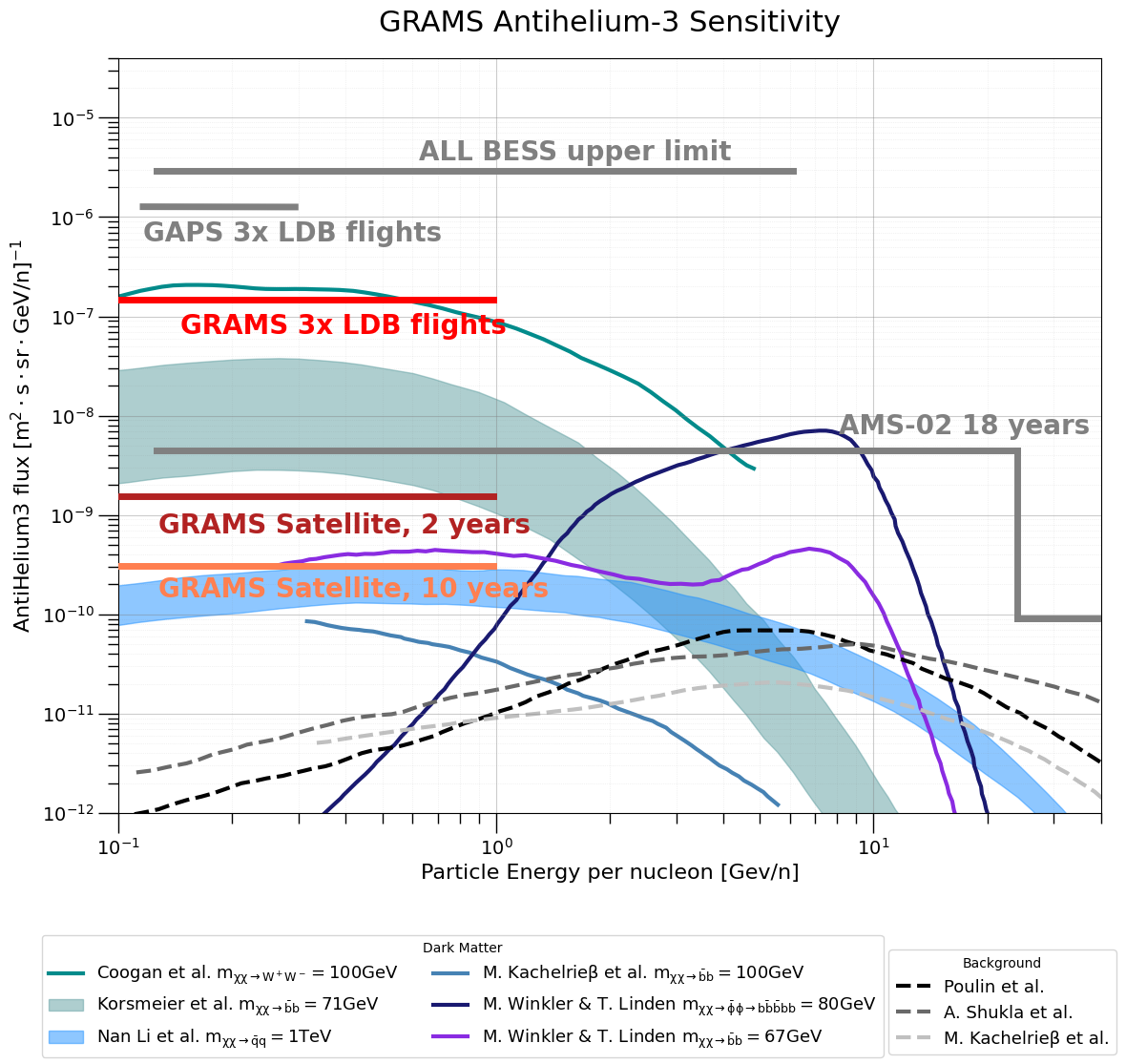}
\end{center}
\caption{The red and brown (coral) lines show the GRAMS antihelium-3 sensitivities for three LDB flights and the satellite missions in comparison with other missions' reported sensitivities  \cite{ saffold2021cosmic}. GRAMS would be able to explore various dark matter models that could produce two orders of magnitude higher antihelium-3 fluxes than standard astrophysical background models \cite{PhysRevD.99.023016, PhysRevD.102.063004, Kachelrie_2020, PhysRevLett.126.101101, PhysRevD.97.103011, PhysRevD.96.083020,  Ding_2019}. The AMS sensitivity and the BESS upper limit to the antihelium flux were estimated based on the $ \overline{He}/He$ ratios \cite{PhysRevLett.108.131301, Kounine:2011bkq} and the precision measurement of the $ He $ flux \cite{PhysRevLett.115.211101, SANUKI2001761} while converting the rigidity to the corresponding kinetic energy for each BESS/AMS rigidity bin.}
\label{antiHe3 sensitivity}
\end{figure}

\begin{table}[htbp]
\centering
\begin{tabular}{l|r}
\hline
&numerical value\\
\hline
$F_{\bar{p}}(E)$ & $10^{-2}$ [m$^2$ s sr GeV/n]$^{-1}$ \\
$F_{^3\overline{He}}^{sec}(E)$ &$5\times10^{-11}$ [m$^2$ s sr GeV/n]$^{-1}$ \\
A&$20\ m\times20\ m$ \\
$\mathbf{\int\Gamma_{\bar{p}}(E)dE}$ & $\mathbf{10.9873\ m^2 sr \ GeV/n}$ \\
$\mathbf{\int\Gamma_{^3\overline{He}}(E)dE}$&$\mathbf{11.6025\ m^2 sr \ GeV/n}$\\
\textbf{T}&\textbf{2 years}/\textbf{10 years}\\
$\mathbf{\varepsilon_g^{\bar{p}}}$&\textbf{$\sim$1}\\
$\mathbf{\varepsilon_g^{^3\overline{He}}}$&\textbf{$\sim$1}\\
$\varepsilon_{TOF}^{\bar{p}}$&0.000142\\
$\varepsilon_{TOF}^{^3\overline{He}}$&0.965\\
$\varepsilon_{\pi}^{\bar{p}}$&0.0000125\\
$\varepsilon_{\pi}^{^3\overline{He}}$&0.912\\
\hline
\end{tabular}
\caption{Sensitivity numerical values for potential satellite mission. Upgraded items marked in bold}
\label{Satellite table}
\end{table}
GRAMS can be expanded to the satellite mission, providing a significantly improved antihelium-3 sensitivity by increasing both the LArTPC detector size and the observation time. Considering the satellite fairing size, the LArTPC can be upgraded to 3.2 m $\times$ 3.2 m $\times$ 0.2 m, which can boost the GRASP for antihelium-3 as seen in Fig.~\ref{GRASP for MAX design}. Here, we assume the observation time to be 2 years/10 years and geomagnetic cutoff $\varepsilon_g\sim1$ at the Lagrange point. With these upgrades listed in table.~\ref{Satellite table}, the GRAMS sensitivity to antihelium-3 nuclei can be significantly improved down to $1.55\times10^{-9}$ [m$^2$ s sr GeV/n]$^{-1}$ / $3.10\times10^{-10}$ [m$^2$ s sr GeV/n]$^{-1}$. Fig.~\ref{antiHe3 sensitivity} shows that GRAMS can uniquely and deeply explore various dark matter models via low-energy antihelium-3 measurements. 

The AMS sensitivity and the BESS upper limit to the antihelium flux were estimated based on the $ \overline{He}/He$ ratios \cite{PhysRevLett.108.131301, Kounine:2011bkq} and the precision measurement of the $ He $ flux \cite{PhysRevLett.115.211101, SANUKI2001761} while converting the rigidity to the corresponding kinetic energy for each BESS/AMS rigidity bin.

\section{Conclusion}
\label{section: Conclusion}

The Low-energy cosmic-ray antihelium-3 nuclei measurement can be a background-free indirect dark matter search method since the antihelium-3 flux from the dark matter annihilation can be a few orders of magnitude higher than the secondary flux from cosmic-ray interactions at the low-energy range. The GRAMS's unique detector and detection concept are optimized to investigate the low-energy range ($E < 1$ GeV/n). We estimated the GRAMS antihelium-3 sensitivity for three LDB flights (105 days of observation time in total) as $1.47\times10^{-7}$ [m$^2$ s sr GeV/n]$^{-1}$. The future GRAMS satellite mission with the upgraded 3.2 m $\times$ 3.2 m $\times$ 0.2 m LArTPC detector and longer observation time (2 years / 10 years), the antihelium-3 sensitivity can be as low as $ 1.55\times10^{-9}$ [m$^2$ s sr GeV/n]$^{-1}$ / $3.10\times10^{-10}$ [m$^2$ s sr GeV/n]$^{-1}$, allowing GRAMS to explore a variety of dark matter models.

\section{Acknowledgments}
This work was supported by the NASA APRA grant, No.22-APRA22-0128 (80NSSC23K1661), and the Alfred P. Sloan Foundation in the US, as well as the Japan Society for the Promotion of Science (JSPS) in Japan. 

\label{section: Acknowledgments}





\bibliographystyle{elsarticle-num}
\bibliography{refs}
\end{document}